\documentclass{jstatphys}
\begin{document}

\title{Comment on ``A Comparison Between Broad Histogram and Multicanonical Methods''}

\author{M.~Kastner%
\footnotemark
\addtocounter{footnote}{-1}
and M.~Promberger%
\footnote{Institut f{\"u}r Theoretische Physik, Friedrich-Alexander-Universit{\"a}t Erlangen-N{\"u}rnberg, Staudtstra{\ss}e 7, 91058 Erlangen, Germany.}
}
\runningauthor{Kastner and Promberger}


\begin{abstract}
A recent paper by A.~R.~Lima, P.~M.~C.~de Oliveira, and T.~J.~P.~Penna [J.~Stat.~Phys.~{\bf 99}:691 (2000)], seems to contain at least two mistakes which deserve comment, one concerning the numerical data, the other being of a conceptual kind.
\end{abstract}

The numerical data of the entropy, presented in Fig.~1 of the above mentioned paper, do not only show statistical errors but also systematic deviations from the exact result. These systematic deviations are apparent in the (physically relevant) {\em gradient} of the entropy, as can be seen by looking for example at the crosses (+) in the inserted plot. The data almost monotonically grow apart from the exact result, which must be interpreted as a signal for something going wrong, either in the simulation itself or in the evaluation of the simulated data. The error will of course in general show up in quantities of interest derived from the entropy. A comparison of a very similar kind---also for a $2d$ Ising system of the same size as in their paper---using correct data without systematic errors can be found in Figs.~3 and 5 of their Ref.~28.\cite{wir}

The second mistake we want to point out is of a conceptual kind. Lima {\em et al.}, although in principle being aware of the independence of the broad histogram method (BHM) from the choice of the stationary distribution (or dynamical rule, to use their own words) underlying the Markov process of the Monte Carlo simulation, somehow fail to present this basic concept properly. An example for this is the inappropriate title of their paper. A Markovian Monte Carlo simulation consists of the following two fundamental parts: The first part is the generation of a sample from configuration space by means of a Markov process according to a particular stationary distribution, for which canonical or multicanonical distributions are examples. The second part is to choose certain observables, for which ``simulation averages'' are calculated from the sample. A histogram is related to a particular choice of such an observable, the BHM to another choice (see, e.g., Ref.~1 for an explicit definition of both these observables). Hence, {\em ``A Comparison Between Broad Histogram and Multicanonical Methods''}, each being related to another of the two {\em independent}\/ basic ingredients of a Monte Carlo simulation, confuses the relevant concepts of both methods and sounds a little like comparing apples with oranges.

The dependence or independence of the BHM or ``other methods'', respectively, on the particular choice of the stationary distribution, also deserves a comment. The authors state that {\em ``...any dynamical rule can be adopted within BHM, the only constraint is to sample with uniform probability the various states belonging to the same energy level...''}.
This restriction is by no means inherent to the BHM. It is merely a consequence of Lima {\em et al.}'s choice to consider simulation averages of quantities like the ``transition rates'' $N$ as functions of the energy only. An arbitrary stationary distribution, depending, for example, on the magnetization, can be chosen, if only the simulation averages of the quantities of interest are recorded as functions of {\em all}\/ parameters on which the stationary distribution depends.%
\footnote{Note that this is {\em not} related to the concept of ``extended transition rates''\cite{wir,multipar} [$N(E,\Delta E,M,\Delta M)$ in a notation following Lima {\em et al.}] monitoring transitions towards states with a certain change of energy $\Delta E$ {\em and} magnetization $\Delta M$. The recording of transitions $N(E,\Delta E,M)$ with a certain $\Delta E$ as a function of energy and magnetization is sufficient to enable the use of a stationary distribution depending on the magnetization. Then, from this quantity, the microcanonical average of $N(E,\Delta E)$ can be obtained.}
Quantities like the density of states or the entropy can be calculated from these data such that the stationary distribution cancels out.


\end{document}